\documentclass[%
 reprint,
superscriptaddress,
 amsmath,
 amssymb,
pra,
floatfix,
]{revtex4-1}

\usepackage{comment}
\usepackage{dsfont}
\usepackage{graphicx}
\usepackage{subcaption}
\usepackage{graphics}
\usepackage{dcolumn}
\usepackage{bm}
\usepackage{float}
\usepackage{qcircuit}
\usepackage{xcolor}

\begin{document}
\preprint{APS/123-QED}

\title{Quantum-classical simulation of two-site dynamical mean-field theory on noisy quantum hardware}

\author{Trevor Keen}
 \affiliation{Department of Physics and Astronomy, University of Tennessee, Knoxville, Tennessee 37996, USA}
 \email{tkeen1@vols.utk.edu}
 
\author{Thomas Maier}
 \affiliation{Computational Sciences and Engineering Division and Center 
 for Nanophase Materials Sciences, Oak Ridge National Laboratory, Oak 
 Ridge, Tennessee 37831, USA}
 
\author{Steven Johnston}
 \affiliation{Department of Physics and Astronomy, University of Tennessee, Knoxville, Tennessee 37996, USA}
 
\author{Pavel Lougovski}%
\affiliation{Quantum Information Science Group, Computational Sciences and 
Engineering Division, Oak Ridge National Laboratory, Oak Ridge, Tennessee 
37831, USA}
 
\date{\today}

\begin{abstract}
We report on a quantum-classical simulation of the single-band Hubbard model using two-site dynamical mean-field theory (DMFT). Our approach uses IBM's superconducting qubit chip to compute the zero-temperature impurity Green's function in the time domain and a classical computer to fit the measured Green's functions and extract their frequency domain parameters. We find that the quantum circuit synthesis (Trotter) and hardware errors lead to incorrect frequency estimates, and subsequently to an inaccurate quasiparticle weight when calculated from the frequency derivative of the self-energy. These errors produce incorrect hybridization parameters that prevent the DMFT algorithm from converging to the correct self-consistent solution. To avoid this pitfall, we compute the quasiparticle weight by integrating the quasiparticle peaks in the spectral function. This method is much less sensitive to Trotter errors and allows the algorithm to converge to self-consistency for a half-filled Mott insulating system after applying quantum error mitigation techniques to the quantum simulation data. 
\end{abstract}

\maketitle


\section{Introduction}

Dynamical mean-field theory (DMFT) is a widely used theoretical framework for modeling strongly correlated electron systems, with specific applications in modeling the Mott transition \cite{Kotliar96}, correlated Hund's metals \cite{HundsReview}, electron-lattice interactions \cite{Werner06, Sangiovanni05}, and advanced electronic structure calculations \cite{Kotliar06}. In simplified terms, DMFT maps the interacting lattice problem onto an impurity problem embedded in a bath of non-interacting electrons, i.e. the Anderson Impurity Model. To accurately approximate the properties of the original lattice model, the embedding is performed self-consistently. This methodology treats the local electronic correlations exactly, while correlations occurring on longer length scales are treated at a mean-field level that retains their dynamics. DMFT becomes exact in the limit in infinite dimensions \cite{Kotliar96}, provided that one can account for the continuum of energy levels constituting the mean-field bath. 

The effectiveness of DMFT is dependent on the impurity solver employed, and several advanced numerical methods have been developed for strongly correlated materials including exact diagonalization (ED) \cite{Liebsch11}, quantum Monte Carlo 
(QMC) \cite{Gull11}, and real-time dynamics with matrix product states (MPS) \cite{Wolf2014MPS}. Each method has its limitations, however. For example, ED approximates the bath with a series of discrete energy levels. It is, 
therefore, limited by the exponential growth of the Hilbert space and can typically handle only a small number of bath levels before exhausting the memory available on a classical computer. QMC is limited by the fermion sign problem, which 
restricts simulations to relatively high temperatures for many models, especially when multiple orbitals are active or when Hund's interactions are included \cite{HundsSign}. 
In comparison to ED, MPS methods suffer less from this exponential memory scaling when using a star geometry for the underlying impurity problem, but suffer from entanglement and normalization issues for other geometries \cite{Wolf2014MPS}.
These examples reflect the broader fact that the classical approaches to exact solutions for strongly correlated systems all suffer from some sort of exponential growth in complexity (e.g. the exponential growth of storage required to store quantum many-body wavefunctions), resulting in an inability to make predictions for larger systems \cite{BES_roundtable}. In a quantum computer, however, the state of the system can be stored and manipulated in qubits. This aspect reduces the simulation problem complexity from exponential in the number of particles to polynomial, giving quantum computers in principle an enormous advantage over classical computers for conducting these simulations. 

\begin{figure}[t]
    \centering
    \includegraphics[width=0.99\linewidth]{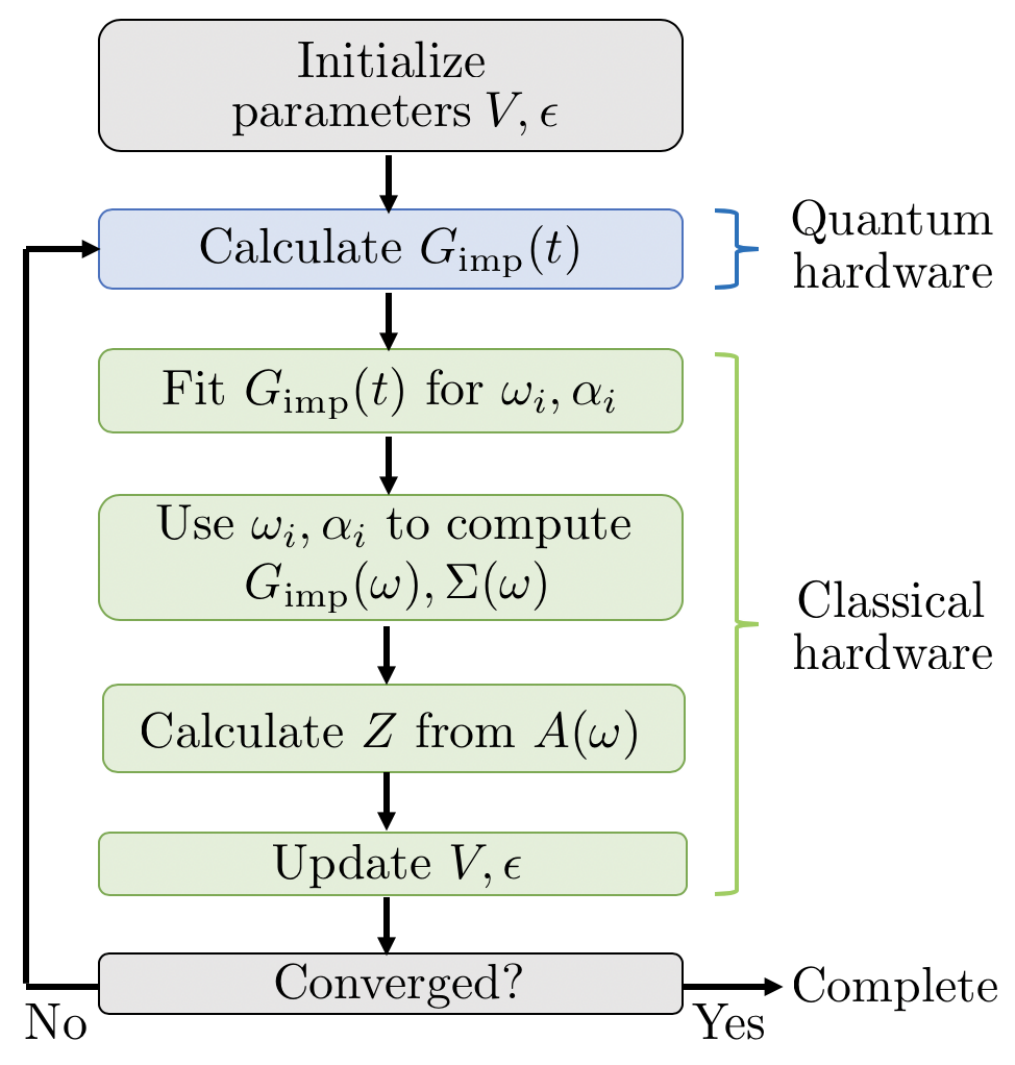}
    \caption{Flowchart for the two-site DMFT calculation implemented on a hybrid classical/quantum system. This loop is repeated until two successive values of $V$ are within some threshold of each other.}
    \label{fig:flowchart}
\end{figure}

In the future, large-scale fault-tolerant quantum computers will enable direct Hamiltonian simulations of many-body systems with thousands of particles. In particular, using quantum computers for strongly correlated electron systems is a valuable and scalable solution as demonstrated by several recent theoretical analyses  (see, e.g. \cite{Wecker2015, Wiebe18, Bauer2016}). In the current era of noisy intermediate-scale quantum (NISQ) \cite{PreskillNISQ} hardware, however, the number of available qubits, their connectivity, and noise prohibit direct implementations of such scalable quantum simulation algorithms. But even with all of their imperfections, NISQ devices can still be leveraged for simulating quantum dynamics in a hybrid quantum-classical algorithmic approach. For example, variational algorithms ~ \cite{McCleanVQE, YuanVQE, Yung2014VQE} use quantum hardware to find expectation values of complex quantum observables such as Hamiltonians while classical computers use those values to update variational parameters in the direction that minimizes the expectation values. DMFT simulations fit naturally into such a hybrid quantum-classical scheme. In the DMFT setting, quantum hardware can be used to  solve the impurity problem which is then post-processed by a classical computer to extract the value of hybridization parameters in a self-consistent manner, see Fig. \ref{fig:flowchart}. Importantly, useful results  that approach the thermodynamic limit can be obtained from DMFT with only a few impurity orbitals \cite{Bauer2016}. 
Moreover, DMFT simulations on a NISQ device are sensible because the impurity is a small part of the lattice. Thus, DMFT will require fewer qubit resources compared to a direct simulation of say, the Hubbard model. It has also been shown that DMFT's limitations, e.g. 
a small set of correlated orbitals and no momentum dependence of the self-energy can be overcome on quantum computers  \cite{Bauer2016}.
Here, we report on an implementation and benchmark of the two-site DMFT scheme described in Ref. \cite{Potthoff2001}. Specifically, we employ one of IBM's superconducting qubit chips to solve the impurity problem 
by measuring the impurity Green's function in the time domain, while the remainder of the DMFT self-consistency loop is executed on a classical computer. For each circuit run on the quantum 
computer, we execute the maximum number of shots allowed by IBM, 8192. We find that the Trotter error associated with the discretization of the time-evolution leads to inaccurate frequency estimates in the fit 
procedure, which in turn introduces an unphysical pole in the self-energy and incorrect  quasiparticle weights. These erroneous frequencies, along with noise from the quantum chip, prevent the DMFT algorithm from 
converging to the correct self-consistent solution. To overcome this issue, we instead determined the quasiparticle weight by integrating the spectral function. We find that this method is much less sensitive to 
gate noise and Trotter error and allows the DMFT algorithm to converge to self-consistency for a half-filled Mott insulator. 

A similar approach to the two-site quantum-classical
DMFT simulation and its implementation on a noiseless quantum simulator was given in Ref. \cite{Kreula2016}. However, only recently have implementations for existing quantum hardware begun to appear \cite{RunggerPreprint}. Though attempting to achieve the same goal -- an implementation of two-site DMFT on a real quantum computer --  our approach differs from that in Ref. \cite{RunggerPreprint} in multiple ways. For one, we apply a Trotterized unitary to directly obtain impurity Green's function data in the
time domain. In contrast, the authors in \cite{RunggerPreprint} use Variational Quantum Eigensolver (VQE)~\cite{McCleanVQE} to implement exact diagonalization. Their method depends on the scalability of the VQE to larger and more complex systems, which is not well known, and these VQE methods are meant to treat Hamiltonians with only a few noncommuting terms~\cite{Bauer2016}. Also, to handle the unphysical poles in the self-energy arising from frequency shifts in the frequency domain representation of the impurity Green's function, the authors of \cite{RunggerPreprint} use a regularization technique to restore the frequency cancellation expected to arise in the Dyson equation. We instead use a different method of calculating quasiparticle weight, which is not explicitly dependent on the self-energy. Another difference is that we iterate the DMFT loop to self-consistency, whereas Ref. \cite{RunggerPreprint} only states that it can be done and did not implement it. 

This paper is organized as follows: Section \ref{ModelandFormalism} introduces the single-band Hubbard Hamiltonian, its mapping onto an Anderson impurity model, and discusses the general DMFT method used to solve the problem. Section \ref{Methods} presents the methods implemented to solve the two-site DMFT problem using a hybrid quantum classical scheme. Our findings are presented in Sec. \ref{Results}. These  include our variational state preparation procedure as well as the fact that Trotter errors and noise lead to an unphysical pole in the self energy, giving incorrect quasiparticle weights, and our method to circumvent this issue. Section \ref{Results} also includes our results for the Mott insulating phase, which were obtained after iterating the hybrid quantum-classical algorithm to self-consistency. Finally, Sec. \ref{Sec:Conclusions} provides some concluding remarks. Appendix A
recounts some of the alternative (unsuccessful) methods we explored to more reliably calculate the quasiparticle weight.

\section{Model \& Formalism} \label{ModelandFormalism}
We implemented a two-site DMFT simulation of the single-band Hubbard Hamiltonian 
\begin{equation} \label{hubbard}
    H = -t \sum\limits_{\langle i, j \rangle, \sigma} \left(c_{i, \sigma}^\dagger c^{\phantom\dagger}_{j, \sigma}+\mathrm{h.c.} \right)
    -\mu\sum_{i,\sigma}\hat{n}_{i,\sigma}
    + U \sum_i \hat{n}_{i, \uparrow} \hat{n}_{i, \downarrow}.
\end{equation}
Here, $\langle \dots \rangle$ denotes a sum over nearest neighbors, 
$c^{\dagger}_{i,\sigma}$ ($c^{\phantom\dagger}_{i,\sigma}$) creates (annihilates) a spin-$\sigma$ ($= \pm \frac{1}{2}$) 
electron on site $i$, $t$ is the nearest-neighbor hopping integral, $\mu$ is the chemical potential, $U$ is the local 
Hubbard repulsion between electrons, and $\hat{n}^{\phantom\dagger}_{i,\sigma} = c^\dagger_{i,\sigma} c^{\phantom\dagger}_{i,\sigma}$ is the number operator. 

The DMFT method maps Eq. (\ref{hubbard}) onto an Anderson impurity model 
\begin{equation} \label{AM}
\begin{split}
    H_\mathrm{AIM} =& \sum\limits_{i=0, \sigma}^{N_\mathrm{bath}} \left(\epsilon^{\phantom\dagger}_i-\mu\right) \hat{n}_{i,\sigma}^{\phantom\dagger} + U\hat{n}_{0,\uparrow}\hat{n}_{0,\downarrow} \\& 
    + \sum\limits_{i = 1, \sigma}^{N_\mathrm{bath}} V^{\phantom\dagger}_i \left( c_{0 \sigma}^{\dagger} 
    c_{i, \sigma}^{\phantom\dagger} + c_{i,\sigma}^{\dagger} c^{\phantom\dagger}_{0, \sigma} \right), 
\end{split}
\end{equation}
where $i = 0$ corresponds to the impurity site and $i=1,\dots,N_\mathrm{bath}$ correspond to the bath sites, $V_i$ is the hybridization 
term that allows hopping between the bath and impurity sites, and $\epsilon_i$ are the bath site energies. 
We consider Eq. (\ref{hubbard}) in infinite dimensions on a Bethe lattice. DMFT is exact in this limit when 
$N_\mathrm{bath}\rightarrow \infty$. 
In what follows, however, we consider the so-called two-site problem with $N_\mathrm{bath} = 1$. While it is a simplified problem, two-site DMFT allows one to recover qualitative results for the Mott transition \cite{Potthoff2001}.

The central quantity in DMFT is the retarded impurity Green's function 
\begin{equation} \label{Gimpr}
    iG_\mathrm{imp}(t) = \theta(t) \langle GS| \{c^{\phantom\dagger}_\sigma (t), c_\sigma^\dagger (0) \} |GS\rangle,
\end{equation}
where $\theta(t)$ is the Heaviside step function, and $|GS\rangle$ denotes the ground state of the system. The impurity Green's function gives the response of the system when a particle is added to or removed from the impurity site. This quantity can be used to compute many useful quantities, e.g. the spectral function and self energy. In the paramagnetic phase, $G_\mathrm{imp}(t)$ is spin symmetric, and so it is sufficient to only compute $G_\mathrm{imp}(t)$ for one spin configuration. 

In the frequency domain $G_\mathrm{imp}(\omega)$ can be expressed as 
\begin{equation} \label{FullGimpw}
    G_\mathrm{imp}( \omega) = \dfrac{1}{\omega + \mu - \Delta(\omega) - \Sigma_\mathrm{imp}(\omega)},
\end{equation}
where $\Delta(\omega) = \frac{V^2}{\omega - (\epsilon_1 - \mu)}$ is the so-called hybridization function that describes the coupling of the impurity to the bath, and $\Sigma_{\mathrm{imp}}$ is the
impurity self-energy. 
In the non-interacting limit ($U=0$), the Green's function reduces to 
\begin{equation} \label{nonintG}
    G^{(0)}_\mathrm{imp} (\omega) = \frac{1}{\omega + \mu - \Delta(\omega)}.
\end{equation} 
The self-energy can be calculated using Eqs.\eqref{FullGimpw} and \eqref{nonintG} together with Dyson's equation 
\begin{equation} \label{dyson}
    \Sigma_\mathrm{imp}^{\phantom{-1}} (\omega) = G_\mathrm{imp}^{(0)} (\omega)^{-1} - G_\mathrm{imp} (\omega)^{-1}.
\end{equation}

We solve this problem for the case of a strong Coulomb repulsion at half-filling, 
where $\epsilon_0-\mu = \frac{U}{2}$ and $\epsilon_1 - \mu = 0$ \cite{Potthoff2001}. This simplification means that we only need to concern ourselves with the self-consistency condition
for the hybridization parameter $V$. 

Equations \eqref{Gimpr}-\eqref{dyson} give the outline of our two-site DMFT protocol, which is also sketched in Fig. \ref{fig:flowchart}. Specifically, we carry out the following 
steps: 
\begin{enumerate}
    \item Fix $U$ and $\epsilon_i - \mu $ to the values 
    appropriate for half-filling, and initialize $V$ to some nonzero initial value.
    \item Measure the impurity Green's function in the time domain.
    \item Fourier transform $iG_{\mathrm{imp}} (t) $ to obtain $G_\mathrm{imp}(\omega)$.
    \item Obtain the spectral function from $G_{\mathrm{imp}} (\omega)$ and the self-energy from $G^{(0)}_{\mathrm{imp}} (\omega) \text{ and } G_{\mathrm{imp}} (\omega)$.
    \item Calculate the quasiparticle weight $\mathcal{Z}$ by integrating the quasiparticle peaks in the spectral function.
    \item Calculate the update to the hybridization parameter $V$ by taking the square root of $\mathcal{Z}$ (this simple square root update method is possible because of the 
    properties of two-site DMFT and the Bethe lattice).
    \item Repeat steps 2-6 with the new value of $V$ until a self-consistent $V$ is reached.
\end{enumerate} 

\section{Methods} \label{Methods}
\subsection{Hardware Needs \& Error Mitigation} \label{hardware_and_mitigation}
Quantum computing simulations of a fermionic system require two qubits for every orbital in the problem, each one to encode the occupancy of the up and down spins on each orbital. Our two-site DMFT protocol will therefore require four qubits. We further require an ancillary qubit to perform a single-qubit interferometry measurement scheme, as described in Refs. 
\cite{Kreula2016, Kreulameas, Dorner2013}, bringing the total number of qubits required to five. We pick a particular subset of qubits on the device that matches the required connectivity to implement our time dynamics circuitry. There is also the circuitry required to prepare the ground state, for which we include the already chosen connectivity between qubits being used for the time dynamics circuitry, and variationally find optimal single qubit rotations between the CNOT gates allowed by connectivity (see Sec. \ref{GSprep} and Fig. \ref{GS_Prep_circ}).

To extract the time dynamics of the impurity Green's function, we implemented the time evolution operator $U(t) = e^{-i H_\mathrm{AIM} t}$ using elementary single and two-qubit gates. There are several approaches that can achieve such a decomposition. We opted to implement this using the first order Trotter-Suzuki expansion as opposed to methods such as qubitization \cite{Low_Qubitization} or the Linear Combinations of Unitary Operations (LCU) \cite{Childs2012, Gui_Lu2006duality, Gui_Lu2009gen_quantu_gates}. While both LCU and qubitization methods achieve a superior scaling in terms of the number of gates needed to implement $U(t)$ for a given $t$ and synthesis error $\epsilon$, we make this choice due to the hardware constraints of current quantum devices. Unlike qubitization and LCU, which require multiple ancillas and the ability to implement advanced controlled unitary operations, Trotterization can be implemented in a more resource-efficient way at the price of increased noise. 
\vspace{-0.05cm}
We also employed several error mitigation techniques to improve our simulations. Specifically, we used the exponential error extrapolation described in Refs. \cite{Suguru2019, Suguru2018} to reduce the noise generated by the relatively large number of CNOT gates required to implement a single Trotter step. We also applied the assignment error reduction method described in the supplementary information of Ref. \cite{Kandala2017} to characterize and correct for qubit readout (assignment) errors.

\subsection{Jordan-Wigner Transformation}
To compute quantities of interest on a quantum computer, we first transformed the fermionic creation and annihilation operators to spin operators  \cite{JW_trans,Bravyi-Kitaev_trans} using the  Jordan-Wigner transformation \cite{JW_trans}.  
In our four qubit system (excluding the ancilla qubit used for measurement), the first two qubits encode the spin-down information for sites one and two, while the third and fourth qubits encode the corresponding  information for the spin-up occupation. We then  represented the creation operator as $\sigma^- = 
X-iY$, following Ref. \cite{Kreula2016}. After applying the Jordan-Wigner transformation, the transformed operators are  
\begin{align} \label{JWtrans}
    \begin{split}
        & c_{1\downarrow}^{\dagger} = \sigma_1^- = \frac{1}{2} \big( 
        X_1 - iY_1 \big), 
        \\& c_{2\downarrow}^{\dagger} = Z_1 \sigma_2^- = \frac{1}{2} 
        Z_1 \big( X_2 - iY_2 \big),
        \\&     c_{1\uparrow}^{\dagger} = Z_1 Z_2 \sigma_3^- =
        \frac{1}{2} Z_1 Z_2 \big( X_3 - iY_3 
        \big),
        \\& c_{2\uparrow}^{\dagger} = Z_1 Z_2 Z_3
        \sigma_4^- = \frac{1}{2} Z_1 Z_2 Z_3
        \big( X_4 - iY_4 \big).
    \end{split}
\end{align}
Here, $X_i$, $Y_i$, or $Z_i$ denote operations where a Pauli operator acts on the $i$\textsuperscript{th} qubit while identity operators act on the remaining qubits. In this representation, the two-site Anderson impurity model is given by 
\begin{widetext}
\begin{equation} \label{JW_HSIAM}
    H_\mathrm{AIM} = \frac{U}{4} ( Z_1 Z_3 - Z_1 -
    Z_3 ) + \frac{\epsilon_0-\mu}{2} (Z_1 +
    Z_3 ) -  \frac{\epsilon_1-\mu}{2} (Z_2 +
    Z_4 )   + \frac{V}{2} ( X_1 X_2 +
    Y_1 Y_2 + X_3 X_4 + Y_3 Y_4 ),
\end{equation}
where we have neglected any identity terms.

\subsection{Trotter Expansion of the time evolution operator}
As mentioned in Sec. \ref{hardware_and_mitigation}, we used a first order Trotter-Suzuki expansion to implement the time evolution operator over higher order methods. The first order Trotter-Suzuki expansion 
\cite{trotter, Suzuki1976} gives
\begin{equation} \label{Trotterized_U}
    \begin{split}
        U(t) =  e^{-i H_{\mathrm{AIM}} t}&\approx  \big( e^{-i \frac{V}{2} (X_1 X_2 + Y_1 Y_2) \Delta t}  
        e^{-i \frac{V}{2} (X_3 X_4 + Y_3 Y_4) \Delta t} e^{-i \frac{U}{4} Z_1 Z_3 \Delta t} 
        \\&~ \times e^{-i (\frac{\epsilon_0 - \mu}{2} - U/4) Z_1 \Delta t} e^{-i (\frac{\epsilon_0 - \mu}{2} - U/4) Z_3 \Delta t} e^{i \frac{\epsilon_1-\mu}{2} Z_2 \Delta t} e^{i \frac{\epsilon_1 - \mu}{2} Z_4 \Delta t} \big)^n + \mathcal{O}(\Delta t^2),
    \end{split}
\end{equation}
where $t$ is the total time, $n$ is the number of time steps taken, and $\Delta t = \frac{t}{n}$. 
In constructing the circuits corresponding to one Trotter step, we utilized the Cartan subalgebra rotation method for each of the $V$ terms \cite{Vidal2004_Cartan_subalg, Coffey2008_Cartan_subalg, Klco2018_Schwinger}, thus reducing CNOT gate costs for the two $V$ terms from six CNOTs each to three CNOTs each. 
\end{widetext}

\subsection{Measurement Scheme and Procedure}
To obtain the values of the impurity Green's function in the time domain, we used a single-qubit interferometry scheme, as proposed in Refs. \cite{Kreula2016, Kreulameas, 
Dorner2013}. We first re-write Eq. \eqref{Gimpr} in terms of
the greater    
$G^>_\mathrm{imp} (t) = -i\langle c_{0\sigma}(t) c_{0\sigma}^\dagger (0) \rangle$ 
and lesser  
$G^<_\mathrm{imp} (t) = i\langle c_{0\sigma}^\dagger (0) c_{0\sigma} (t) \rangle$ 
Green's functions.
We then use the Jordan-Wigner Transformation [Eq. \eqref{JWtrans}] to recast these as
\begin{equation} \label{JWGgreater}
    \begin{split}
        G^>_\mathrm{imp} (t) = &\frac{-i}{4} \big[ \langle U^\dagger (t) X_1 U (t) X_1 \rangle 
        -i \langle U^\dagger (t) X_1 U (t) Y_1 \rangle
        \\ & +i \langle U^\dagger (t) Y_1 U (t) X_1 \rangle
        + \langle U^\dagger (t) Y_1 U (t) Y_1 \rangle \big]
    \end{split}
\end{equation}
and
\begin{equation} \label{JWGlesser}
    \begin{split}
        G^<_\mathrm{imp} (t) = &\frac{i}{4} \big[ \langle X_1 U^\dagger (t) X_1 U (t) \rangle
        + i \langle X_1 U^\dagger (t) Y_1 U (t) \rangle
        \\& - i \langle Y_1 U^\dagger (t) X_1 U (t) \rangle
            +   \langle Y_1 U^\dagger (t) Y_1 U (t) \rangle \big].
    \end{split}
\end{equation}

After measuring the retarded impurity Green's function $G_\mathrm{imp}(t)$ at each Trotter step, we least-squares fit $i G_\mathrm{imp}(t)$ on a classical computer using the the scipy package \cite{scipy} and a function of the form 
\begin{equation} \label{fit}
    i G_\mathrm{imp} (t) = 2\left[\alpha_1 \cos(\omega_1 t) + \alpha_2 \cos(\omega_2 t)\right], 
\end{equation}
which is a simplification due to the assumed particle-hole symmetry in our system \cite{Kreula2016}. The 
Fourier transform of Eq. \eqref{fit} is straightforward with  
\begin{equation} \label{Gimpw}
    \begin{split}
        G_\mathrm{imp}(\omega + i\delta) & =  \alpha_1 \left( \dfrac{1}{\omega + i\delta +\omega_1} + \dfrac{1}{\omega + i\delta - \omega_1} 
        \right) 
        \\& + \alpha_2 \left( \dfrac{1}{\omega + i\delta +\omega_2} + \dfrac{1}{\omega + i\delta - \omega_2} \right), 
    \end{split}
\end{equation}
where $\delta$ is an artificial broadening. Once self-consistency is reached and the fit parameters are obtained, we use the Dyson equation [Eq. \eqref{dyson}] to compute the self-energy and, subsequently, the spectral function $A(\omega) = -\frac{1}{\pi}\text{Im} [ G_\mathrm{imp} (\omega+i\delta) ]$.

\section{Results} \label{Results}
\subsection{Ground State Preparation} \label{GSprep}
The main obstacle for performing fermionic calculations on a quantum computer lies in preparing the necessary eigenstates. The quantum phase estimation algorithm \cite{Nielsen_Chuang} will not work for the hardware we have used due to the inability to feed forward the state acquired via phase estimation to the time dynamics part of the algorithm. Instead, we use a variational approach that is well-suited to the limited connectivity of IBM's quantum chips.
\begin{figure}[h]
    \centering
    \[\Qcircuit @C=1em @R=.3em {
    |0\rangle & &  \gate{R_y(\theta_1)} & \qw & \qw & \ctrl{2} & \gate{R_y(\theta_7)} & \ctrl{1} & \qw 
    \\
    |0\rangle & &  \gate{R_y(\theta_2)}  & \qw & \qw & \qw  & \qw & \targ & \qw \\
    |0\rangle & &  \gate{R_y(\theta_3)} & \ctrl{1} & \gate{R_y(\theta_5)} & \targ & \gate{R_y(\theta_8)} & \qw & \qw \\
    |0\rangle & &  \gate{R_y(\theta_4)} & \targ & \gate{R_y(\theta_6)} & \qw & \qw & \qw & \qw \\
    }\]   
    \caption{The circuit used to prepare the ground state using only three CNOT gates and eight single qubit rotations. The parameters \{$\theta_i$\} are varied to maximize the fidelity between the output state of this circuit and the ground state of the system.}
    \label{GS_Prep_circ}
\end{figure}
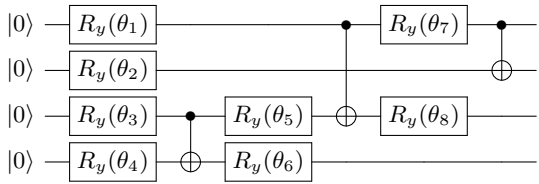
Our variational state ansatz can be prepared by a shallow circuit with three CNOTs and eight single-qubit rotations (see Fig. \ref{GS_Prep_circ} for details). The single-qubit rotation parameters are chosen to minimize the expectation value of the Hamiltonian $H_\mathrm{AIM}$ for given values of $V,U,\epsilon_{i},\mu$. We find that this ansatz can reproduce the exact ground state (to the precision of the minimization).
More specifically, our variational state has a fidelity with the exact ground state of 1 with an error on the order of $10^{-14}$. When $V$ becomes smaller than $10^{-2}$, we neglect the $V$ term and can prepare the ground state exactly.

\begin{figure} 
\centering
 \includegraphics[width=0.96\columnwidth]{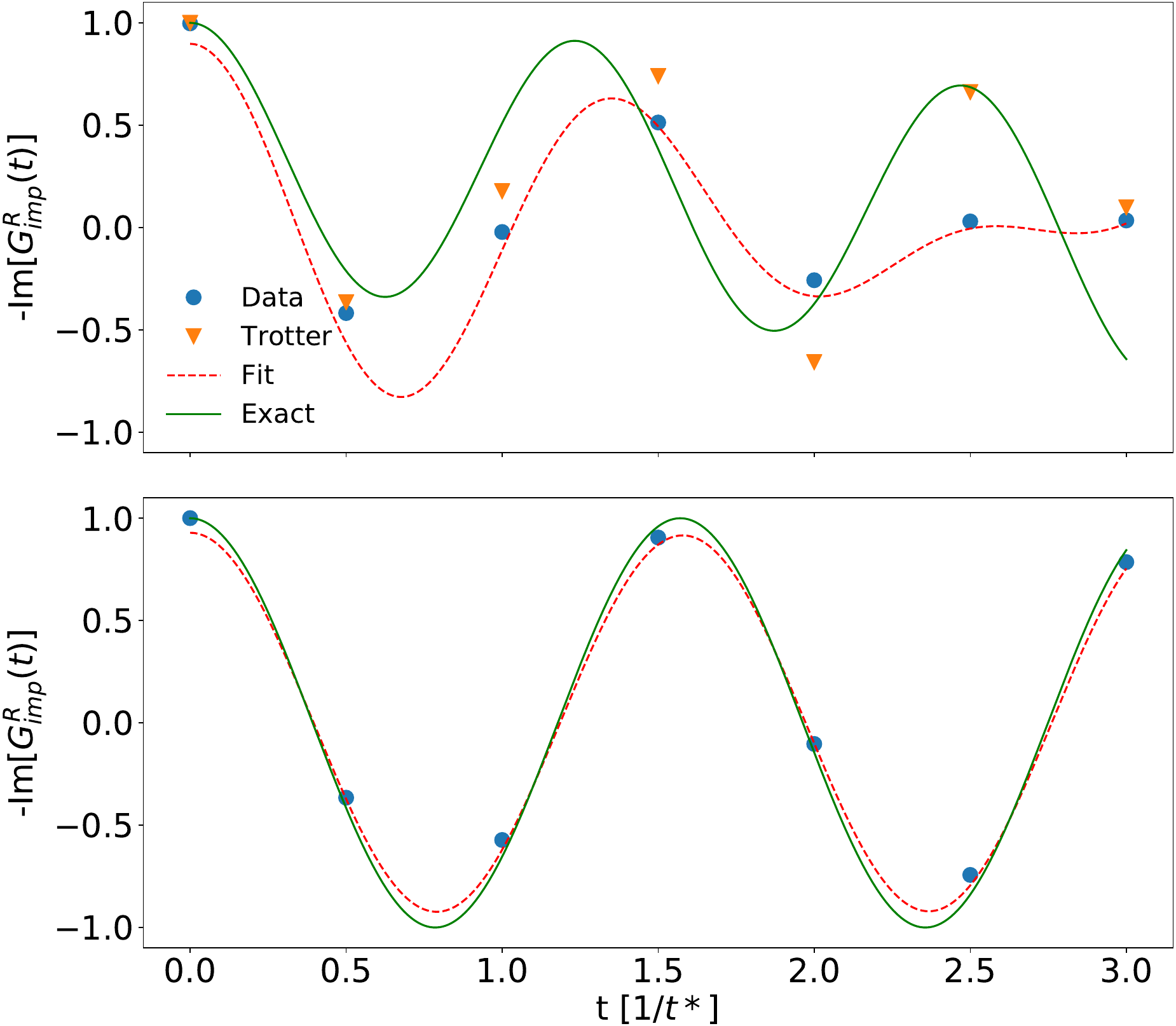}
 \caption{Top: Data and fit for the impurity Green's function at the first step in the self-consistency loop with $U=8t^* \text{ and } V=t^*$ compared against the exact result and the result with Trotter error only. The parameters for the fit shown are $\omega_1 = 4.033$, $\omega_2 = 5.197$, $\alpha_1 = 0.242$, and $\alpha_2 = 0.207$. Bottom: Data and fit for impurity Green's function at self-consistency $V=0$ with $U=8t^*$ plotted along with the exact result. The parameters for the fit shown are $\omega_1 = 3.980$, $\omega_2 = 2.116$, $\alpha_1 = 0.461$, and $\alpha_2 = 0.003$. Note that in the bottom plot, the calculation with Trotter error only is absent since there is no error from Trotterization when $V=0$.}
 \label{Gimps}
\end{figure}

\subsection{Impurity Green's Function}
\begin{figure}[t] 
  \centering
  \includegraphics[width=0.99\linewidth]{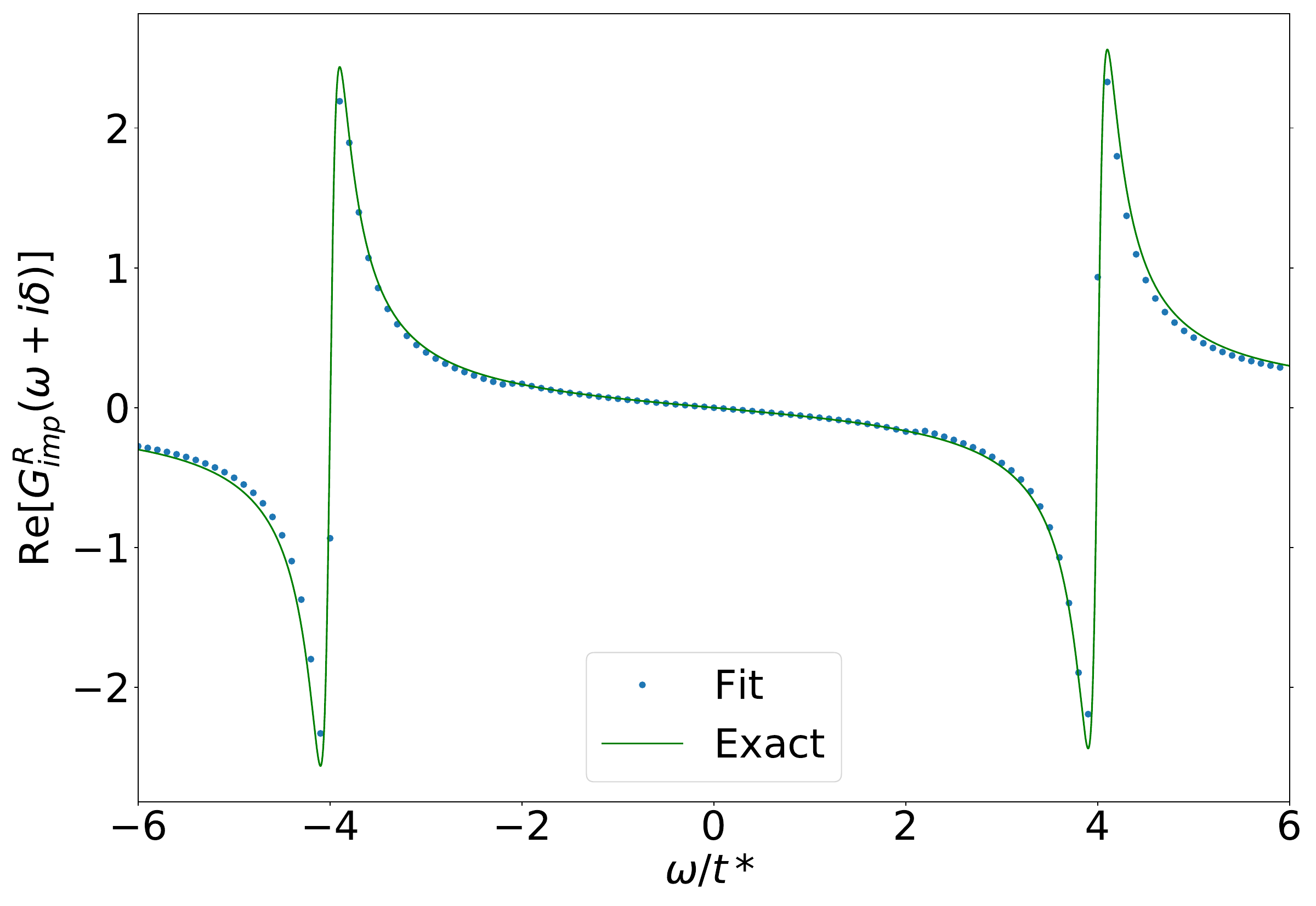}
  \caption{Impurity Green's function in the frequency domain for $U = 8t^*$, here calculated via Eq. \eqref{Gimpw} after the DMFT algorithm has converged to self-consistency. The data are compared to the exact result, and both curves assume a broadening of $\delta=0.1$.}
  \label{Gimpw_fig}
\end{figure}

\begin{figure}[t] 
\centering
\includegraphics[width=0.99\linewidth]{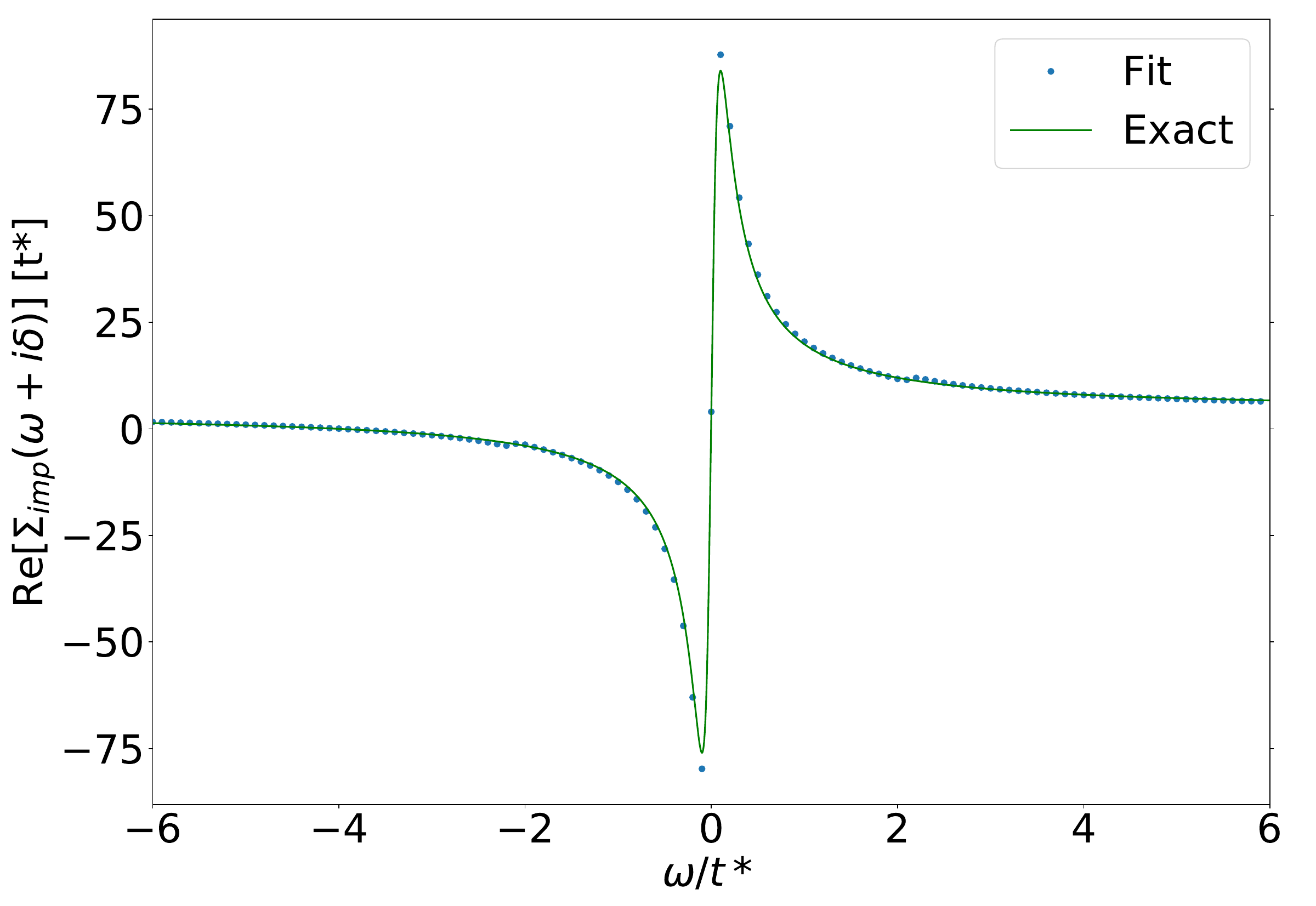}
\caption{The real part of the self-energy calculated from Fig. \ref{Gimpw_fig} via Eq. \eqref{dyson}. Data are shown for the fit parameters and the exact result, both with a broadening of $\delta=0.1$.}
\label{self_fig}
\end{figure}

As stated previously, the impurity Green's function is the central quantity of interest in the DMFT routine. In Fig. 3, we show the impurity Green's function in the time domain for two different sets of 
parameters, namely $V=t^*$ (top) and $V=0$ (bottom) with $U=8t^*$ for both cases. The data in Fig. \ref{Gimps} are superimposed with the fits to the data [Eq. \eqref{fit}] and the exact solution for those 
parameters. In the top panel of Fig. 3 we also plot the impurity Green's function calculated with only the error introduced by the Suzuki-Trotter approximation to the Green's function. This curve is absent in the bottom plot since the Trotter error is zero for $V=0$ and so those data points would lie directly on top of the exact curve. In both cases, there are only seven data points for  $G_\mathrm{imp}(t)$ because the Trotter step is so expensive in terms of CNOT gates that the noise generated for more time steps and a nonzero $V$ 
would overwhelm the simulation. In Fig.  \ref{Gimpw_fig}, we show the impurity Green's function in the frequency domain extracted from the fit parameters [Eq. \eqref{Gimpw}], along with the exact solution, both obtained after 
self-consistency is achieved ($V=0$). In Fig. \ref{self_fig}, we display the self-energy of the system at self-consistency, calculated using Eq. \eqref{dyson} with the $G_{\mathrm{imp}} (\omega)$ shown in Fig. 
\ref{Gimpw_fig}. As expected for two-site DMFT at half-filling with $U > U_c = 6t^*$, at self-consistency one term in Eq. \eqref{fit} dominates with a frequency at $\frac{U}{2}$. Due to noise,
however, our self-consistent solution does not converge to exactly the right frequency (it is shifted by approximately 0.02/$t^*$). Nevertheless, we still obtain good agreement with the exact solution.

\subsection{Quasiparticle Weight Calculations}
Because of the semicircular form for the density of states of the Bethe lattice in the limit of infinite coordination, the hopping parameter $V$ in the case of a single bath level is given simply by the square root of the quasiparticle weight $V = \sqrt{\mathcal{Z}}$ \cite{Potthoff2001}. The latter can be calculated from the self-energy using the relation 
\begin{equation} \label{quasipart_weight}
    \mathcal{Z}^{-1} = 1-\frac{d \mathrm{Re}[\Sigma(\omega)]}{d\omega}\bigg|_{\omega = 0}\,.
\end{equation}
In practice, however, we found that the Trotter error and noise inherent to the quantum simulation result in slight shifts in the fit frequencies $\omega_1$ and $\omega_2$ [see Eq. (\ref{fit})]. These errors produce extraneous peaks around $\omega=0$ in the self-energy computed using the Dyson equation, which gives small nonzero quasiparticle weights, regardless of the other parameters. 
We observed that even small errors in the frequencies due to Trotterization causes unreliable derivatives and thus unreliable  quasiparticle weights. This issue can be mitigated by taking more Trotter steps, but with the noise restrictions of the available quantum computers, we are restricted to approximately six Trotter steps.

\begin{figure}[t] 
    \centering
    \includegraphics[width=\columnwidth]{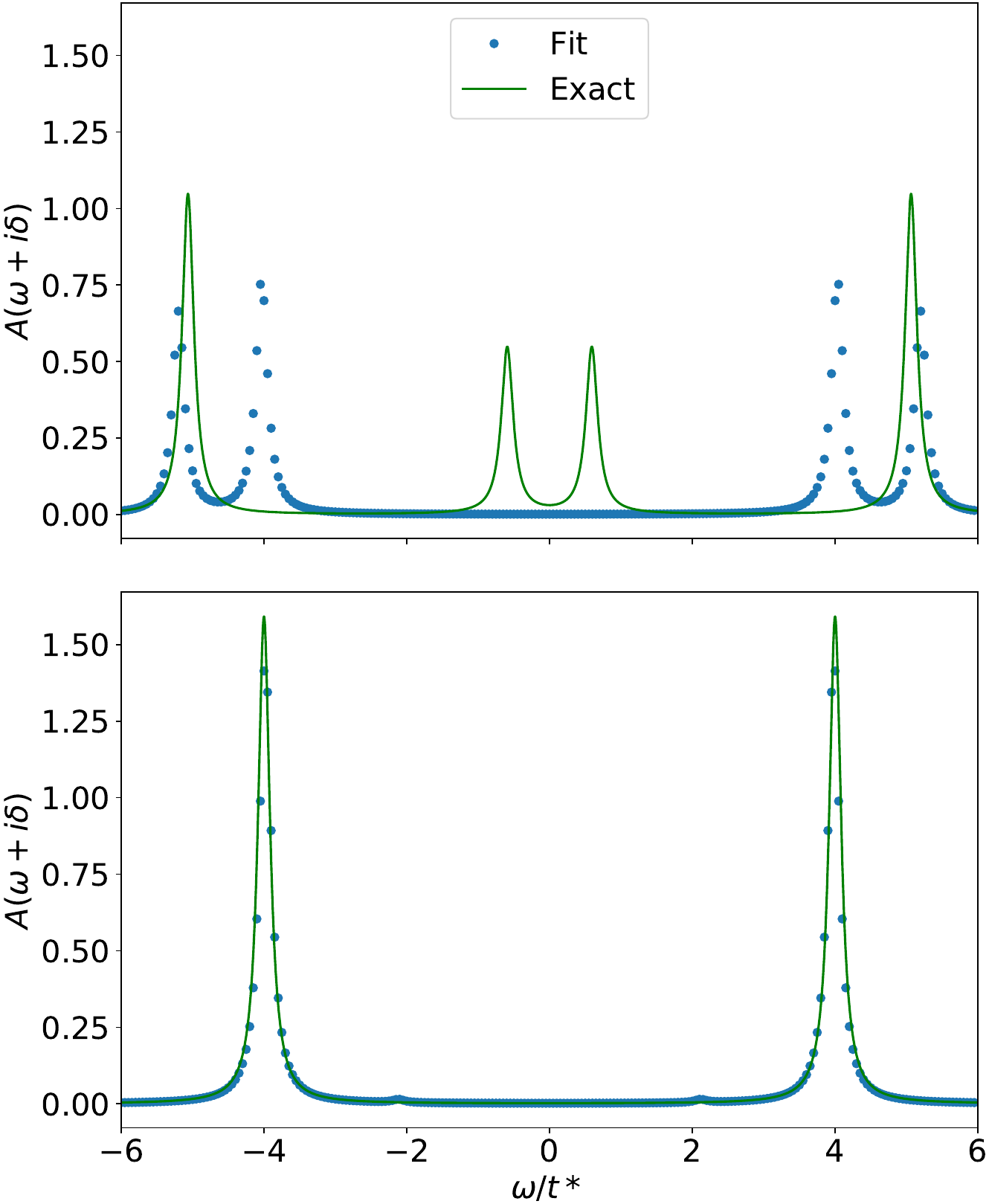}
    \caption{Top: The calculated spectral function after the first step of the self-consistency loop 
    with $U=8t^*$, $V=1t^*$, and a broadening of $\delta=0.1$, compared with the exact result. 
    Bottom: The same spectral function after the DMFT loop has converged to $V=0$.}
    \label{spec_fig} 
\end{figure}

To circumvent this issue, we instead integrate the quasiparticle peaks, i.e. the two peaks closest to $\omega=0$, in the spectral function to obtain the quasiparticle weight. For example, in the top 
panel in Fig. \ref{spec_fig}, the two innermost peaks of the spectral function are visible for finite $V$, but for our Mott insulating case at self consistency they become very small. This method still produces inaccurate quasiparticle weights, but they are less sensitive to the shifts in frequency due to Trotter error, and accurate enough to allow us to obtain some meaningful results. 

\begin{figure}[t]
    \centering
    \includegraphics[width=\columnwidth, keepaspectratio]{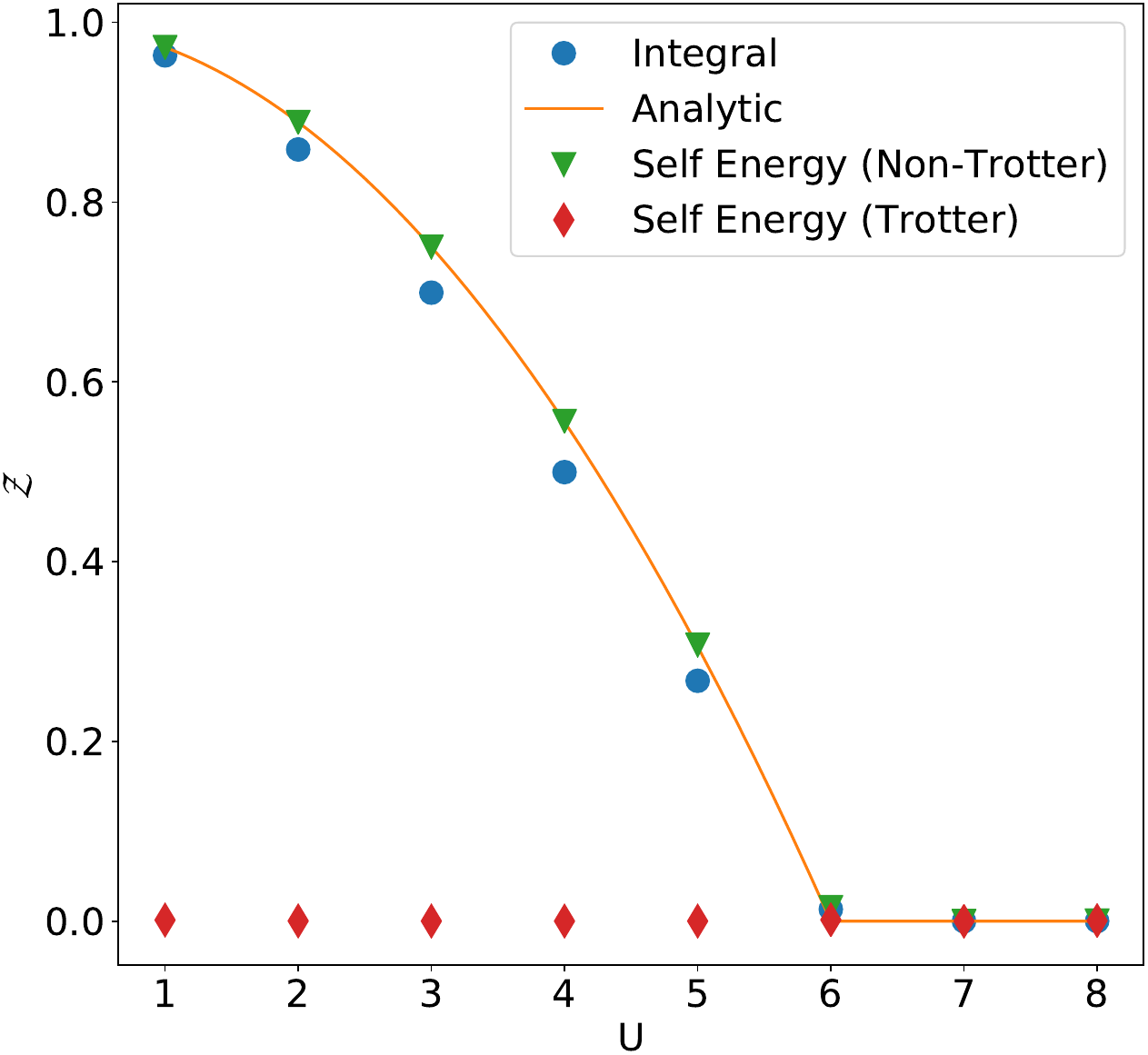}
    \caption{Quasiparticle weight at self-consistency as a function of $U$ using Eq. \eqref{quasipart_weight} with fit parameters from both the Trotterized unitary (diamonds), the exact unitary fit parameters (triangles), and from integrating the 
    low-energy peaks of the spectral function (circles) with fit parameters from the Trotterized unitary, 
    along with the analytical result of Ref. \cite{Potthoff2001} (solid line).}
    \label{fig:ZvsU}
\end{figure}

For finite values of $V$, the fitting procedure gives incorrect parameters when the data for $iG_{\mathrm{imp}}(t)$ is fit to Eq. \eqref{fit} due to the limited number of Trotter steps that we can implement, and the noise inherent to current quantum hardware. 
These erroneous fit parameters make the updates for the self-consistency parameters inaccurate. Because of this, we have found it difficult to  
converge to self-consistency when $U<U_c$ and a metallic solution ($V \ne 0$) is expected. In Fig. \ref{fig:ZvsU}, we show the values of the quasiparticle weight at self-consistency for different values of $U$. We see that with Trotter error, the values of the quasiparticle weight calculated via Eq. \eqref{quasipart_weight} are completely unreliable. Fig. \ref{fig:ZvsU} also shows that we do not recover the exact quasiparticle weight at self-consistency for all values of $U$, but obtain fairly good results that are more resilient to Trotter error in comparison to any other method we attempted (see appendix \ref{appendixA}), and that for our trial case of the strongly Mott insulating regime, we can recover the exact quasiparticle weight at self-consistency. It should be noted that all of the data in Fig. \ref{fig:ZvsU} was calculated on a classical computer.We are, however, able to obtain a converged solution 
for $U > U_c$, where a Mott insulating gap forms and at self-consistency $V = 0$, as discussed in the next section. Other methods that we attempted to employ to calculate 
the quasiparticle weight more reliably are given in Appendix A.

\subsection{Mott Insulating Phase}
For an on-site impurity Coulomb repulsion above a critical value of $U_c = 
6t^*$ at half-filling ($\epsilon_0-\mu = \frac{U}{2}$ and $\epsilon_1 - \mu = 0$), the self-consistent value of $V$ is zero. This solution corresponds to the well known Mott insulating phase \cite{Potthoff2001}. In 
our particular case, we set $U=8t^*$ and took an initial guess for the hybridization parameter of $V=1t^*$, see the top figure in Fig. \ref{Gimps} for the initial run. We then iterated our approach to the self-consistent $V=0$ solution, with the condition that once $V$ is sufficiently small ($V\leq10^{-2}$), we neglected the $V$ term and solve what is essentially the single site problem. 
The bottom panel in Fig. \ref{Gimps}, Figs. \ref{Gimpw_fig} and \ref{self_fig}, and the bottom panel in Fig. \ref{spec_fig} show the resulting impurity Green's functions, self-energy, and spectral functions, respectively, obtained once the DMFT loop has converged. 
This regime gives poles for the impurity Green's function at $\pm \frac{U}{2}$. 
Although there is no Trotter error at self-consistency for this case, noise from the quantum computer gives a small but finite value for the amplitude $\alpha_2$ of the second cosine in Eq. \eqref{fit}, even though the exact solution has $\alpha_2 = 0$.
This error is the origin of the small peaks located near $\omega/t^* \approx \pm 2$ in the bottom panel of Fig. \ref{spec_fig}.
Nevertheless, our results demonstrate that the DMFT loop for the two-site problem can be iterated to convergence for parameters in the Mott insulating regime. 

\subsection{Trotter Error Analysis}
As mentioned previously, we found that the Trotter error accumulated after several Trotter steps implemented on a quantum computer results in shifted frequencies obtained from the fit. This error causes a mismatch between the poles in $G_\mathrm{imp}^{(0)} (\omega) \text{ and } G_\mathrm{imp}$, leading to unphysical poles in the self-energy. The noise introduced by the quantum computer will exacerbate this issue. This result agrees with the findings of Ref. \cite{RunggerPreprint}. For a Trotterized unitary such that 
\begin{equation} \label{TrotError}
   \vert\vert U-U_T \vert\vert \leq \delta_T,
\end{equation}
where $U$ is the full unitary, $U_T$ is the Trotterized unitary, and $\delta_T$ is the Trotter error. For our case, we find that the Trotter error
incurred in both $G^>_\mathrm{imp}$ and $G^<_\mathrm{imp}$ is less than or equal to $2\delta_T$. For our first order Trotter expansion,
and our relatively large time step ($\Delta t = 0.5$) required to satisfy the Nyquist criteria with a reasonable number of Trotter steps, this Trotter error is significant.

\section{Conclusions}\label{Sec:Conclusions}
We have implemented an algorithm to conduct the two-site dynamical mean-field theory calculations on a quantum computer, employing multiple error mitigation strategies. Due to limited connectivity of the IBM superconducting qubit quantum computers, we use a variational ansatz to prepare the ground state of the system, greatly reducing the cost in terms of CNOT gates. We found that Trotter error and noise lead to frequencies shifted from their true values, which in turn lead to an unphysical pole in the self-energy. These aspects lead to unreliable calculations for the quasiparticle weight, and the update of the impurity-bath hybridization parameter $V$. These limitations prevented the DMFT algorithm from reaching self-consistency.  
To overcome this problem, we integrated the quasiparticle peaks in the spectral function to obtain updates to the hybridization parameter. Using this alternative method, we were able to iterate the DMFT loop to 
self-consistency for a strong-coupling Mott insulating phase. We were, however, unable to obtain self-consistency in the metallic phase.

Our work highlights several of the challenges in implementing quantum many body algorithms on NISQ devices. For example, to go beyond two-site DMFT with currently available quantum
computing hardware, other methods will need to be employed for calculating the Green's functions, such as those proposed in 
\cite{endo2019calculation_green's, kosugi2019construction_green's}, 
or a more complex version of the regularization proposed in Ref. \cite{RunggerPreprint}.

\begin{acknowledgements}
This work was performed in part at Oak Ridge National Laboratory,  operated by UT-Battelle for the U.S. Department of Energy under Contract No. DE-AC05-00OR22725. This work is supported by the U.S. Department of Energy, Office of Science, Office of Advanced Scientific Computing Research (ASCR) Quantum Algorithm Teams (QAT) and Quantum Computing Application Teams (QCATS) programs, under field work proposal numbers ERKJ333 and ERKJ347. S.~J.~acknowledges additional support the National Science Foundation under Grant No. MPS-1937008. This research used quantum computing system resources supported by the U.S. Department of Energy, Office of Science, Office of Advanced Scientific Computing Research program office. Oak Ridge National Laboratory manages access to the IBM Q System as part of the IBM Q Network. The views expressed are those of the authors and do not reflect the official policy or position of IBM or the IBM Q team.
\end{acknowledgements}

\appendix 
\section{Different Methods of Calculating Quasiparticle Weight} \label{appendixA}
A possible alternative to the proposed methods for calculating the quasiparticle weight is to use the Kramers-Kronig relations between the real and imaginary parts of the self-energy to relate $\frac{d \mathrm{Re}[\Sigma(\omega)]}{d\omega}\big|_{\omega = 0}$ to an  
integral over the imaginary part of the self-energy. This method may be preferable since in many cases the ``quasiparticle peaks" in the spectral function may not be as pronounced and/or well separated from the rest of the spectrum as here. 
The integration over the entire spectral range should make this method less sensitive to the unphysical near zero frequency structure in the self-energy, but it is not expected to be entirely immune to this problem. For our case, we found this Kramers-Kronig based 
method for calculating the derivative of $\frac{d \mathrm{Re}[\Sigma(\omega)]}{d\omega}\big|_{\omega = 0}$ to be more accurate than directly taking the derivative on the real axis, but less accurate than integrating the quasiparticle peak of the spectral function for the number of Trotter steps implementable on available quantum computers. 

In another attempt to mitigate the errors in calculating the quasiparticle weight, we introduced a small fictitious temperature and transformed all of our quantities
to the Matsubara frequency domain. Specifically, we performed the Hilbert transform of Eq. \eqref{Gimpw} to obtain the Green function in terms of Matsubara frequency. From this,
we obtained the self-energy at the first Matsubara frequency as a function of (ficticious) temperature from the Dyson equation. From these quantities, we obtained the imaginary frequency quasiparticle weight as a function of temperature
$$
\mathcal{Z}(T) = \dfrac{1}{1-\dfrac{\mathrm{Im}[\Sigma(\pi T)]}{\pi T}}
$$

which becomes identical to the real frequency quasiparticle weight in the zero temperature limit. We calculated ${\cal Z}(T)$ for many small fictitious temperatures and extrapolated to zero temperature.
\begin{figure}[H] 
    \centering
    \includegraphics[width=\columnwidth]{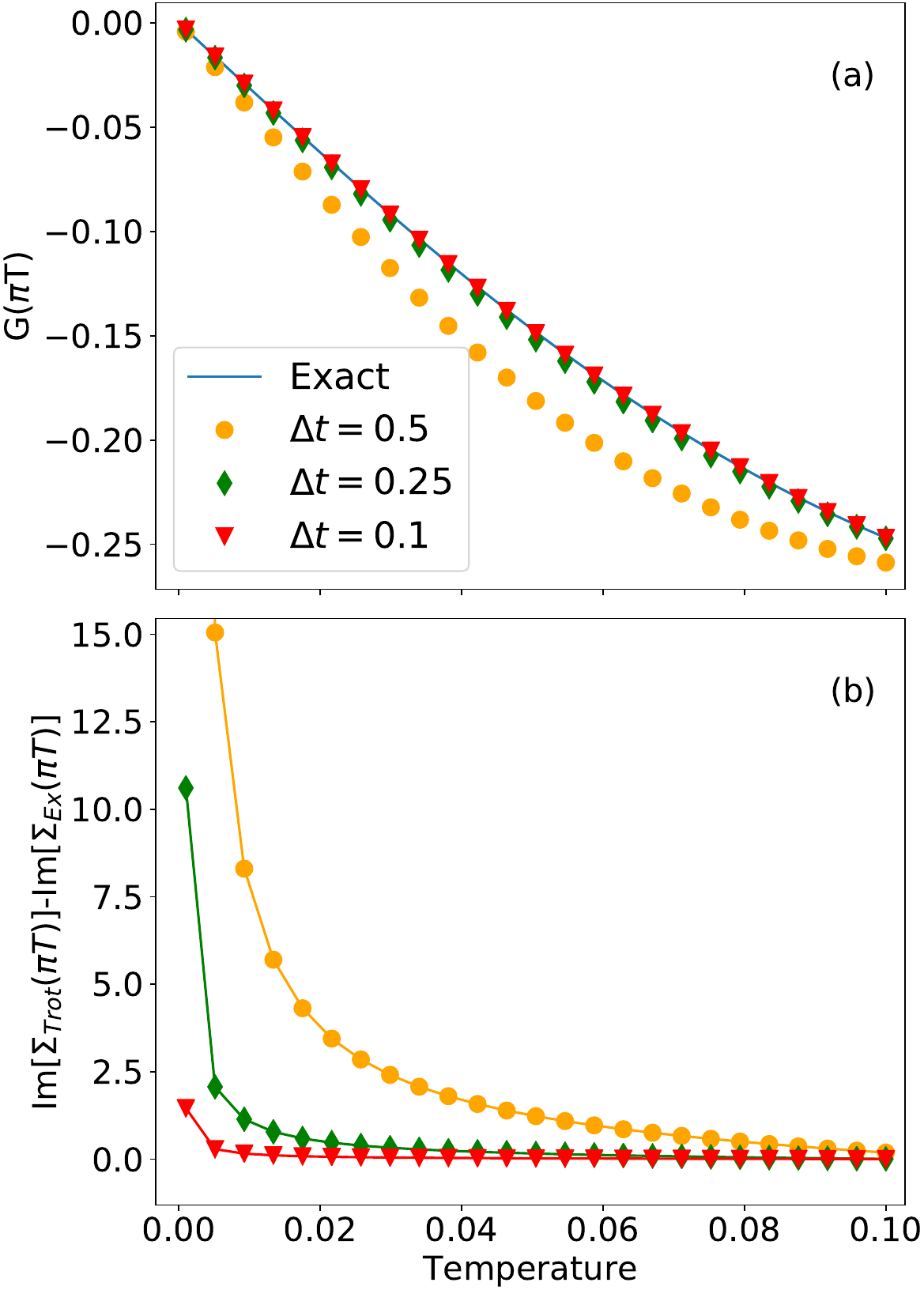}
    \caption{(a) Matsubara Green function at the first Matsubara frequency vs. temperature for different Trotter step sizes at $U=8t^*$ and $V=t^*$.
             (b) Difference between the self-energy computed with Trotter fit parameters and the exact self-energy at the first Matsubara frequency vs. temperature for different Trotter step sizes at $U=8t^*$ and $V=t^*$.}
    \label{Matsubara_fig} 
\end{figure}
We again found that the Trotter error caused this method to give completely unreliable results for a Trotter step size of more than a few thousandths, making this method completely impractical for near-term applications. Figure \ref{Matsubara_fig}(a) shows the Matsubara Green function
at the first Matsubara frequency vs. temperature for different size Trotter steps. Figure \ref{Matsubara_fig}(b) shows the difference between the Matsubara self-energy with no Trotter error
and the Matsubara self-energy with different Trotter step sizes vs. temperature, with both self-energies being evaluated at the first Matsubara frequency.

While the Green's function appears to converge rapidly with decreasing Trotter step size, the non-linear relation between the self-energy and the Green's function leads to a large error in the self-energy even for Trotter step sizes where the Green's function is very close to the exact result. 

\bibliography{references}

\end{document}